\documentclass[11pt,conference,onecolumn]{IEEEtran}

\usepackage{amsmath,amssymb,graphicx}

\DeclareMathOperator*{\argmax}{arg\,max}

\newcommand{\bx}{\ensuremath{\boldsymbol{x}}}
\newcommand{\by}{\ensuremath{\boldsymbol{y}}}
\newcommand{\blambda}{\ensuremath{\boldsymbol{\lambda}}}
\newcommand{\bgamma}{\ensuremath{\boldsymbol{\gamma}}}
\newcommand{\bphi}{\ensuremath{\boldsymbol{\phi}}}
\newcommand{\bmu}{\ensuremath{\boldsymbol{\mu}}}

\newcommand{\bt}{\ensuremath{\boldsymbol{t}}}
\newcommand{\bxi}{\ensuremath{\boldsymbol{x_i}}}

\newcommand{\bxj}{\ensuremath{\boldsymbol{x_j}}}

\newcommand{\bF}{\ensuremath{\boldsymbol{f}}}
\newcommand{\bFi}{\ensuremath{\boldsymbol{f_i}}}
\newcommand{\bFj}{\ensuremath{\boldsymbol{f_j}}}
\newcommand{\bxin}{\ensuremath{\boldsymbol{X_i}}}
\newcommand{\bxjn}{\ensuremath{\boldsymbol{X_j}}}
\newcommand{\byn}{\ensuremath{\boldsymbol{Y}}}
\newcommand{\bxn}{\ensuremath{\boldsymbol{X}}}

\newcommand{\bFcap}{\ensuremath{\boldsymbol{F}}}

\newcommand{\cd}{\ensuremath{\mathcal{D}}}

\newcommand{\cn}{\ensuremath{\mathcal{N}}}

\newcommand{\cs}{\ensuremath{\mathcal{S}}}

\newcommand{\cx}{\ensuremath{\mathcal{X}}}
\newcommand{\cy}{\ensuremath{\mathcal{Y}}}

\newtheorem{theorem}{Theorem}

\begin{document}

\title{Sensing Capacity for Markov Random Fields}


\author{\authorblockN{Yaron Rachlin,
Rohit Negi, and Pradeep Khosla}
\authorblockA{Department of Electrical and Computer Engineering, Carnegie Mellon University\\ 5000 Forbes Ave., Pittsburgh, PA
15213, Email: \{rachlin, negi, pkk\}@ece.cmu.edu}}

%

\maketitle

\begin{abstract}
This paper computes the sensing capacity of a sensor network, with
sensors of limited range, sensing a two-dimensional Markov random
field, by modeling the sensing operation as an encoder. Sensor
observations are dependent across sensors, and the sensor network
output across different states of the environment is neither
identically nor independently distributed. Using a random coding
argument, based on the theory of types, we prove a lower bound on
the sensing capacity of the network, which characterizes the
ability of the sensor network to distinguish among environments
with Markov structure, to within a desired accuracy.
\end{abstract}

\section{Introduction} \label{sec:introduction}
We investigate how spatial Markov structure in the environment
affects the number of sensors required to sense that environment
to within a desired accuracy. We explore this relationship in the
context of discrete sensor network applications such as
distributed detection and classification. The number of sensors
required to achieve a desired performance level depends the
characteristics of the environment (e.g. target sparsity, likely
target configurations, target contiguity), the constituent sensors
(e.g. noise, range, sensing function), and the resource
constraints at sensor nodes (e.g. power, computation,
communications). Resource constraints such as communications and
power are important to consider in the design of sensor networks
due to the limitations they impose on, among other things, network
lifetime and sampling rate. See, for example,
\cite{DuarteMelo03,Scaglione02,Gatspar04} for a discussion on the
effects of resource constraints on sensor networks. However, even
if these resource constraints were eliminated, many basic
questions about the theoretical design limitations of sensor
networks are not yet adequately addressed. The sensing
capabilities of the sensors, the spatial characteristics of the
environment being sensed, and the required accuracy of the sensing
task impose sharp limitations on the number of sensors required to
achieve a desired performance level. We elucidate this purely
sensing-based limitation, by demonstrating a lower bound on the
minimum number of sensors required to achieve a desired sensing
performance, given the sensing capabilities of the sensors and a
spatial Markov model of the environment. External constraints,
such as power, communication, bandwidth, and computation are not
considered in this paper.

\begin{figure}[t]
\begin{center}
\includegraphics[height=1.0cm]{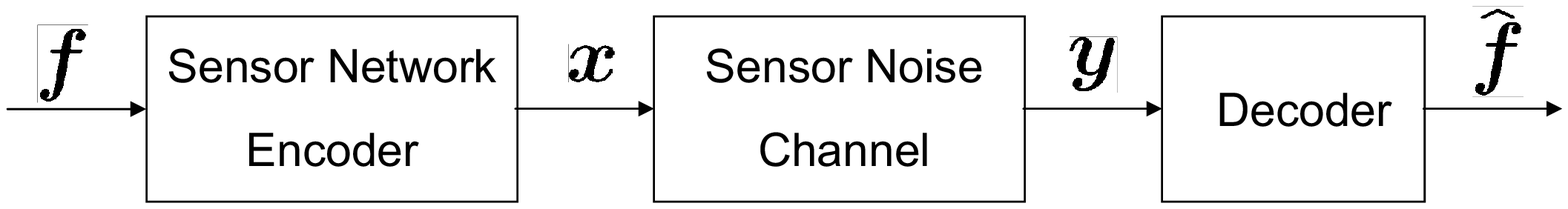}
\caption{Sensor network modeled as a channel encoder.}
\label{channelmodel}
\end{center}
\end{figure}

We model the presence/absence of targets in a two-dimensional grid
as a Markov random field \cite{Li01},
 and the sensor network as a `channel
encoder' (Figure \ref{channelmodel}). This `encoder' maps the grid
of targets into a vector of sensor outputs, which corresponds to a
``codeword.'' These sensor outputs are then corrupted by noise.
The decoder observes this noisy codeword and provides an estimate
of the spatial target configuration. Viewing the sensor network as
a channel encoder allows us to use ideas from Shannon coding
theory. However the messages do not necessarily occur with equal
probability, unlike messages in classical channel codes. In
addition, as we will show, the ``codebook'' obtained has codewords
which are neither independent nor identical. These differences
require a novel analysis and a novel concept of `sensing capacity'
$C(D)$. The distortion $D$ is the maximum tolerable fraction of
spatial positions which may be erroneously sensed. For a given
$D$, $C(D)$ represents the maximum ratio of the total number of
target positions under observation to the number of sensors, such
that below this ratio, there exist sensor networks whose average
probability of error goes to zero as the number of possible target
positions and sensors goes to infinity.

In previous work \cite{RachlinITW04}, we introduced the concept of
a sensing capacity. We extended this work in \cite{RachlinIPSN05}
to account for arbitrary sensing functions and localized sensing
of a one-dimensional target vector, with i.i.d. targets. In this
paper we explore the effect of Markov structure in a
two-dimensional environment on the sensing capacity, as occurs in
several practical applications (e.g. robotic demining and
prospecting, distributed surveillance). We model the environment
as a Markov random field, and show an extension of the theory of
types to include Markov random fields. Section \ref{sec:model}
introduces and motivates our sensor network model. Section
\ref{sec:theorem} states a lower bound on sensing capacity for the
model. Illustrative calculations of the sensing capacity appear in
Section \ref{sec:sim}.

\section{Sensor Network Model} \label{sec:model}

We denote random variables and functions by upper-case letters,
and instantiations or constants by lower-case letters. Bold-font
denotes vectors. $\log(\cdot)$ has base-2. Sets are denoted using
calligraphic script. $D(P||Q)$ denotes the Kullback-Leibler
distance and $H(P)$ denotes entropy.

We consider the problem of sensing discrete two dimensional
environments with spatial structure. Examples include camera
networks that localize people in a room, seismic sensor networks
that localize moving objects, minefield mapping, and soil mapping.
There exists a large body of work in distributed detection
\cite{Varshney97}, but we are not aware of the existence of any
`sensing capacity' results. \cite{Chakrabarty01} introduced the
idea of viewing sensor networks as encoders, and used algebraic
coding theory to design highly structured sensor networks, but no
notion of capacity was discussed.



\begin{figure}[t]
\begin{center}
\includegraphics[height=3.5cm]{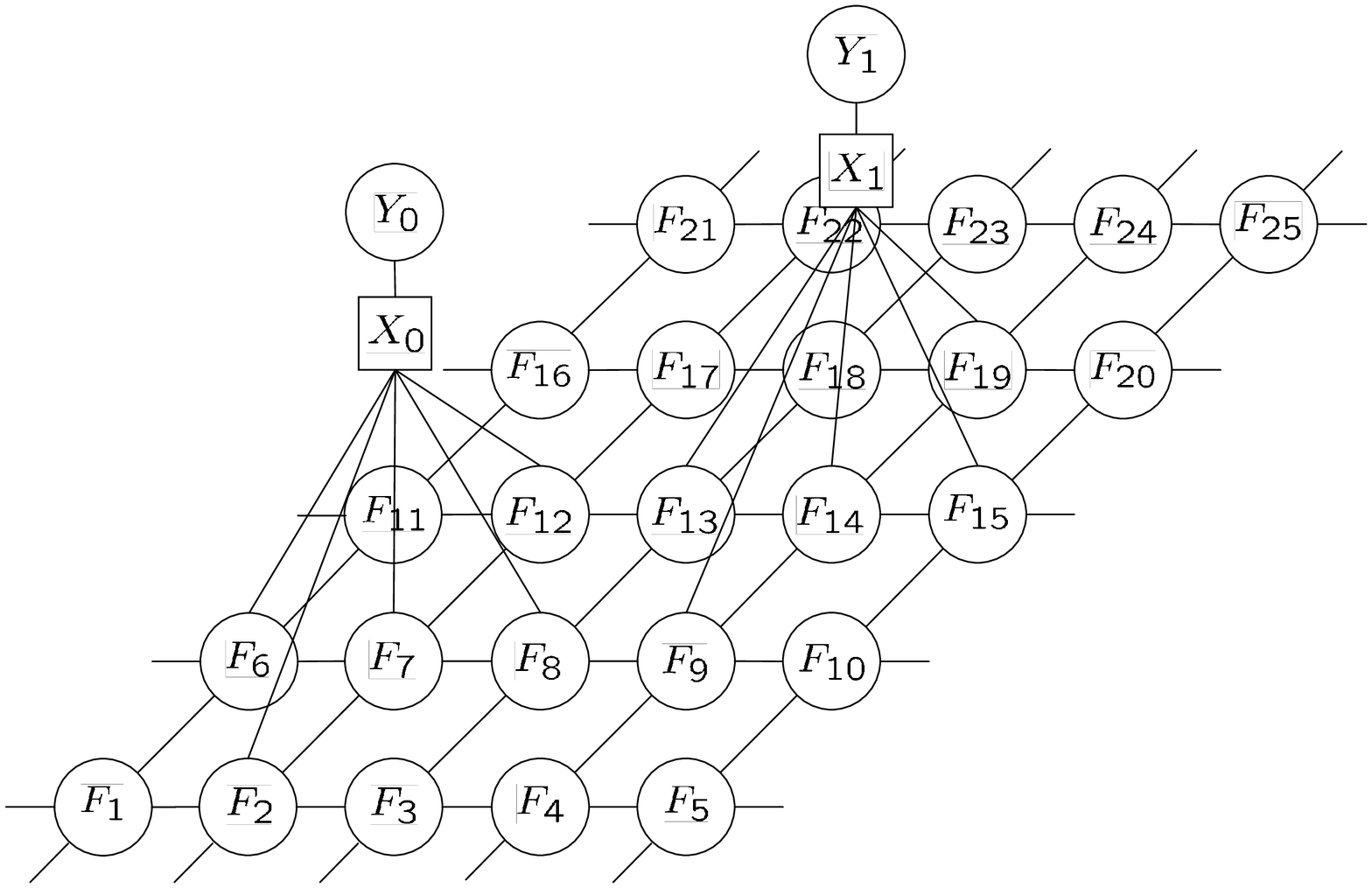}
\caption{Sensor network model with $k=5, n=2, c=1$.}
\label{sensornetworkmodel}
\end{center}
\vspace{-0.3in}
\end{figure}

The model we present attempts to abstractly characterize
 the discrete sensor network applications listed above. Figure
\ref{sensornetworkmodel} shows an example of our sensor network
model. There are $k^2$ discrete spatial positions that need to be
sensed in a $k\times k$ grid. Each discrete position may contain
no target or one target, though extensions to non-binary targets
is straightforward. Thus, the target configuration is represented
by a $k^2$-bit `target field' $\bF$. The possible target fields
are denoted $\bFi$, $i\in\{1,\ldots,2^{k^2}\}$. We say that `a
certain $\bF$ has occurred' if that field represents the true
spatial target configuration.  Target fields occur with
probability $P_{\bFcap}(\bF)$ and are assumed to be distributed as
a pairwise Markov random field (also referred to as an auto-model)
\cite{Li01}, a widely used model (e.g. distributed detection,
image processing) that captures spatial dependencies while still
allowing for efficient algorithms. This model differs from the
equiprobable i.i.d. target distribution explored in our previous
work, and allows one to model environments with structure such as
target sparsity, likely target configurations, and spatial
contiguity among targets. We remark that the methods used in this
paper can be directly extended to more complex Markov field models
(besides pairwise Markov), at the price of more cumbersome
notation. A pairwise Markov random field (Figure
\ref{sensornetworkmodel}) is modeled as a graph, where each target
position $F_{h}$ corresponds to a node. The subscript $h$ indexes
the set of possible grid locations. The set of four grid blocks
directly adjacent to $F_{h}$, which are neighbors of $F_{h}$ in
the graph, are written as $\cn_{h}$. We assume circular boundary
conditions; i.e. the targets on the boundaries are adjacent to the
opposite boundary. We assume that all $\bF$ have positive
probability, and that given its neighbors, the probability of a
target is independent of the remaining targets. According to the
Hammersley-Clifford theorem, a Markov random field that obeys
these two properties is distributed as a Gibbs distribution
\cite{Li01}. A Gibbs distribution is written as a normalized
product of positive functions over the cliques in the graph of the
Markov random field. In our pairwise Markov random field there are
two types of cliques: single nodes $\{F_{h}\}$ with associated
function $\frac{1}{W}P_F$, and pairwise cliques
$\{(F_{h},F_{v}):v\in \cn_h\}$ with associated function
$P_{F|F'}$. The constant $W$ is defined as
$W=\sum_{\bt\in\{0,1\}^5}P_{F}(t_5)\prod_{r=1}^{4}
P_{F|F'}(t_5|t_r)$. Thus, we have the following Gibbs distribution
for $\bF$,
\begin{eqnarray}
P_{\bFcap}(\bF)= Z^{-1}
\prod_{h}W^{-1}P_F(f_{h})\prod_{v\in\cn_{h}}P_{F|F'}(f_{h}|f_{v})
\label{mrfeqn}
\end{eqnarray}
where $Z$ is a normalization constant.

The sensor network has $n$ identical sensors. Sensor $\ell$
located at grid block $F_h$ senses (i.e. is connected to in the
graph) a set of contiguous target positions within a Euclidean
distance $c$ of its location (though our approach can be extended
to other sensor coverage models). Circular boundary conditions,
discussed above, are assumed.
Figure \ref{sensornetworkmodel} depicts sensors with range $c=1$.
Each sensor outputs a value $x \in {\cal X}$ that is an arbitrary
function of the targets which it senses, $x=\Psi(\{f_{v}:
v\in\cs_{c,h}\})$, where $\cs_{c,h}$ is the coverage of a sensor
located at grid block $F_h$ with range $c$. Since the number of
targets sensed by a target depends only on the sensor range, we
write the number of targets in a sensor's coverage as $|\cs_c|$.
One example of a sensing function is a weighted sum of the
targets. This function corresponds to a seismic sensor, which
senses the weighted sum of target vibrations. The `ideal output
vector' of the sensor network $\bx$ depends on the sensor
connections, sensing function, and on the target field $\bF$.
However, we assume that each sensor output $y \in\mathcal{Y}$ is
corrupted by noise, so that the conditional p.m.f. $P_{Y|X}(y|x)$
determines the observed output. Since the sensors are identical,
$P_{Y|X}$ is the same for all the sensors. Further, we assume that
the noise is independent in the sensors, so that the `sensor
output vector' $\by$ relates to the ideal output $\bx$ as
$P_{\byn|\bxn}(\by|\bx)=\prod_{\ell=1}^{n}P_{Y|X}(y_{\ell}|x_{\ell})$.
Observing the output $\by$, a decoder (described below) must
determine which of the $2^{k^2}$ target fields $\bFi$ occurred.

We define the sensor network $S(k^2,n,c)$ as a graph (Figure
\ref{sensornetworkmodel}) with connections between $n$ sensors and
the $k^2$ spatial positions, and the noise corrupted observations
of the ideal sensor outputs. We assume a simple model for randomly
constructing such sensor networks, where each sensor chooses a
region of Euclidean radius $c$ (as constructed above) with equal
probability among the set of possible regions of radius $c$. This
would occur, for example, if sensors were randomly dropped on a
field, or robots moved randomly over a region.

\section{Sensor Network Capacity Theorem} \label{sec:theorem}

For a sensor network, randomly generated as explained above, the
ideal output $\bx$ is a function of the sensor network
instantiation $S(k^2,n,c)$, the sensing function $\Psi$, and the
occurring target field $\bF$. Denote $\bxin$ as the random vector
which occurs when $\bFi$ is the target field (where $\bxin$ is
random because of the random generation of the sensor network
$S(k^2,n,c)$). Since each sensor independently forms connections
to a subset of targets, $P_{\bxin}(\bxi)=\prod_{\ell=1}^{n}
P_{X_i}(x_{i\ell})$. However, it is important to note that when
not conditioned on the occurrence of a specific target field
$\bFi$, the sensor outputs are not independent. Further, we also
note that the random vectors $\bxin$ and $\bxjn$, associated with
a \emph{pair of target fields} $\bFi$ and $\bFj$ respectively, are
\emph{not independent}, since the sensor network configuration
produces a dependency between them (i.e. similar target fields are
likely to produce a similar sensor network output). Thus, the
`codewords' $\{\bxin, i=1,2,\ldots,2^{k^2}\}$ of the sensor
network (one corresponding to each $\bFi$) are non-identical and
dependent on each other, unlike channel codes in classical
information theory. Further the messages \{$\bFi$\} to which these
'codewords' correspond are not equally likely, necessitating a
different analysis.

Given the noise corrupted sensor network output $\by$, we estimate
the target field $\bF$ which generated this noisy output by using
a decoder $g(\by)$. We allow the decoder a distortion of
$d\in[0,1]$. Given $D_{\text{H}}(\bFi,\bFj)$ is the Hamming
distance between two target fields, given that the tolerable
distortion region of $\bFi$ is
$\cd_i=\{j:\frac{1}{k^2}D_{\text{H}}(\bFi,\bFj)<d\}$, and given
that $\bFi$ occurred, the probability of error is
$P_{e,i,s}=\text{Pr}[\text{error}|i,s,\bxi, \by ]=\text{Pr}[g(\by)
\not\in\cd_i|i,s,\bxi,\by ]$. Averaging $P_{e,i,s}$ over all
sensor networks,
 we write
the average error probability, given $\bFi$ occurred,
 as $P_{e,i}=E[P_{e,i,s}]$.
We use average error probability $P_{e}= \sum_i \
P_{e,i}P_{\bFcap}(\bFi)$ as our error metric.

We define the `rate' of the sensor network as the ratio of target
positions to sensors, $R=\frac{k^2}{n}$. Given a tolerable
distortion $D$, we call $R$ achievable if the sequence of sensors
networks $S(\lceil nR\rceil,n,c)$ satisfies $P_{e} \rightarrow 0$
as $n\rightarrow\infty$. The \emph{sensing capacity}
 of the sensor network is defined as
$C(D)=\max R$  over achievable $R$.

The main result of this paper is to show that the sensing capacity
$C(D)$ of the sensor network model presented in this paper is
non-zero, and to characterize it as a function of environmental
structure $P_{\bFcap}$, noise $P_{Y|X}$, sensing function $\Psi$,
and sensor range $c$. The proof broadly follows the proof of
channel capacity provided by Gallager \cite{Gallager68}, by
analyzing a union bound of pair-wise error probabilities, averaged
over randomly generated sensor networks. However, it differs from
\cite{Gallager68} in several important ways. One primary
difference arises due to our `encoder' (i.e. sensor network).
Rather than randomly generating pairwise independent codewords as
in the Shannon capacity proof, our encoder corresponds to a
randomly generated sensor network. Given this encoder (sensor
network), the codewords are dependent on each other and
non-identically distributed. To overcome this complication, we
observe that since each sensor in our network randomly chooses a
set of contiguous targets, we can use the method of types
\cite{Csiszar98} to group the exponential number of pair-wise
error probability terms into a polynomial number of terms in order
to prove convergence of error probability. A second primary
difference is that we analyze two-dimensional messages that are
not equally likely. Thus, rather than using a maximum likelihood
decoder in our proof we use a maximum a posteriori decoder.
Further, the statement of the main result requires the extension
of the existing definition of higher order types \cite{Csiszar98}
to accommodate two-dimensional fields. In our proof, we will use
two kinds of types.

The field type $\bphi$: Since the probability distribution of a
pairwise Markov random field has a factorized form, depending only
on quintuplets of values as shown in (\ref{mrfeqn}), we can
rewrite the probability of a Markov random field as a product over
the set of possible quintuplets. Each term in the product will
have a degree equal to the number of times that quintuplet of
values occurred in the field. We refer to the vector of normalized
counts of the number of times each quintuplet occurred in a field
$\bF_i$ as the field type $\bphi_i$. $\bphi_i$ is a normalized
thirty-two dimensional vector for binary fields. (\ref{mrfeqn})
can be rewritten in terms of $\bphi$ as follows,
\begin{equation}
P_{\bFcap}(\bF)= \frac{1}{Z}\prod_{\{\bt\} \in \{0,1\}^5}
2^{k^2\phi_{\bt}\log(\frac{1}{W}P_F(t_5)\prod_{r=1}^{4}
P_{F|F'}(t_5|t_r))} \label{mrftype}
\end{equation}

The sensor types $\bgamma$ and $\blambda$: For a sensor located
randomly in the target field, the probability of a sensor
producing a value depends on the number of target patterns that
correspond to the sensor's range, and thus, can be written as a
function of the frequency with which each pattern occurs in the
field. The sensor type $\bgamma_i$ is a vector that corresponds to
the normalized counts over the set of possible target
configurations in the sensor's field of view in a field $\bF_i$.
For a sensor of range $c$, $\bgamma_i$ is a $2^{|\cs_c|}$
dimensional vector, where each entry in the vector $\bgamma_i$
corresponds to the frequency of occurrence of one of the possible
$|\cs_c|$ bit patterns.

Since each sensor independently chooses a set of contiguous
spatial positions to sense, the distribution of its ideal output
$X_{i}$ (which is sensed when the $i^{th}$ target field $\bFi$
occurs) depends only on the type $\bgamma$ of $\bFi$. i.e., for a
sensing function $\Psi$, a range $c$, and a target field $\bF_i$
of type $\bgamma_i$, $P_{\bxin}(\bxi)= P^{\bgamma_{i,n}}(\bxi) =
\prod_{\ell=1}^{n} P^{\bgamma_i }(x_{i\ell})$ for all $\bFi$ of
type $\bgamma_i$ \cite{RachlinITW04}.

Next, we note that for sensor of range $c$ the conditional
probability $P_{\bxjn|\bxin}$ depends on the \emph{joint sensor
type} $\blambda$ of the $i^{th}$ and $j^{th}$ target fields
$\bFi,\bFj$. $\blambda$ is the matrix of $\lambda_{(t_1\ldots
t_{|\cs_c|})(u_1\ldots u_{|\cs_c|})}$, the fraction of positions
in $\bFi,\bFj$ where $\bFi$ has a target pattern $t_1\ldots
t_{|\cs_c|}$ while $\bFj$ has a target pattern $u_1\ldots
u_{|\cs_c|}$. We denote the set of all joint sensor types for
sensors of range $c$ observing a target field of area $k^2$, as
$\Lambda_{k^2}(c)$. Since the output of each sensor depends only
on the contiguous region of targets which it senses,
 $P_{\bxjn|\bxin}$ depends only on $\blambda$ \cite{RachlinITW04}. Thus, $P_{\bxjn|\bxin}(\bxj | \bxi) =
\prod_{\ell=1}^{n} P^{\blambda}(x_{j\ell} | x_{i\ell})$ for all
$i,j$ of the same joint type $\blambda$.

The field types $\bphi$ and the sensor types $\bgamma$ of a field
$\bF$ must be consistent with each other. Due to the circular
boundary conditions of our Markov random field graph, the
marginals of types are precisely equal to types over smaller sets.
Thus when $c>1$, $\bphi$ can be obtained precisely by
marginalizing $\bgamma$, while for $c=0$ $\bgamma$ can be obtained
by marginalizing $\bphi$. For $c=1$ the two types are identical.
Further, $\blambda$ also allows computation of $\lambda_{(1)(0)}$
and $\lambda_{(0)(1)}$. These latter quantities correspond to the
number of grid locations where field $i$ has a target and field
$j$ does not, and vice versa.

We specify two probability distributions which we will utilize in
the main theorem. The first is the joint distribution of the ideal
output $\bxi$ when $\bFi$ occurs and the noise corrupted version
$\by$ of $\bxi$. i.e., $P_{\bxin\byn}(\bxi,\by)=
\prod_{\ell=1}^{n}P_{X_iY}(x_{i\ell},y_{\ell}) =
\prod_{\ell=1}^{n}P_{X_i}(x_{i\ell})P_{Y|X}(y_{\ell}|x_{i\ell})$.
The second distribution  is the joint distribution of the ideal
output $\bxi$ corresponding to $\bFi$
 and the noise corrupted output
$\by$ generated by the occurrence of a \emph{different} target
field $\bFj$. We can write this joint distribution as
$Q_{\bxin\byn}^{(j)}(\bxi,\by)=
\prod_{\ell=1}^{n}Q_{X_iY}^{(j)}(x_{i\ell},y_{\ell})=
\prod_{\ell=1}^{n}\sum_{a \in {\cx}}P_{X_i}(x_{i\ell})
P_{X_j|X_i}(x_j=a|x_{i\ell})P_{Y|X}(y_{\ell}|x_j=a)$. Note that
$\bxin,\byn$ are dependent here, although $\byn$ was produced by
$\bxjn$ because of the dependence of $\bxin,\bxjn$. This is unlike
Shannon codes, where the codewords are independent.

Since each sensor in the sensor network depends only on the
targets in the contiguous spatial region which it observes,
$P_{\bxin\byn}(\bxi,\by)$ depends only on the sensor type
$\bgamma$ of $\bFi$. Thus, we write \;\;\;
$P_{\bxin\byn}(\bxi,\by) = \prod_{\ell=1}^{n}
P_{{X_{i}Y}}^{\bgamma}(x_{i\ell},y_{\ell})$ where
$P_{{X_{i}Y}}^{\bgamma}(x_{i},y) =
P^{\bgamma}(x_{i})P_{Y|X}(y|x_{i})$. Similarly,
$Q_{\bxin\byn}^{(j)}(\bxi,\by)$ depends only on the joint sensor
type $\blambda$ of $\bFi,\bFj$ and can be written as
\;$\prod_{\ell=1}^{n} Q_{{X_{i}Y}}^{\blambda}(x_{i\ell},y_{\ell})$
where $Q_{{X_{i}Y}}^{\blambda}(x_{i},y) = \sum_{a \in {\cx}}
P^{\bgamma}(x_{i}) P^{\blambda}(x_j=a|x_{i}) P_{Y|X}(y|x_j=a)$. We
are now ready to state the main theorem of this paper.

\begin{theorem}[Sensing Capacity for pairwise MRF, $c\geq1$] The
sensing capacity at distortion $D$ for target field distribution
$P_{\bFcap}$ satisfies,
\begin{equation} \label{theorem}
\hspace{-1em}C(D) \geq C_{LB}(D) =
 \min_{\bgamma_i \in T(\bphi^{\ast})} \min_{\begin{subarray}{c}\blambda \\
\lambda_{(0)(1)}+\lambda_{(1)(0)}>D
\end{subarray}}\frac{D\left(P_{X_iY}^{\bgamma_i}\|Q_{{X_{i}Y}}^{\blambda}\right)}{DENOM}
\end{equation}
where $DENOM =
H(\blambda)-H(\bgamma_i)+H(\bphi^{\ast})-D(\bphi_j\|\frac{1}{W}P_F\prod_{r=1}^{4}P_{F|F'})-H(\bphi_j))$,
where the sensors have range $c\geq1$, and where
${\bgamma_{i}},{\bgamma_{j}}$ are obtained by marginalizing
$\blambda \in \Lambda_{k^2}(c)$. Here, $T(\bphi^{\ast})$ consists
of the set of $\bgamma$ that marginalize to the typical
$\bphi^{\ast}$ (the $\bphi_i$ such that
$D(\bphi_i\|\frac{1}{W}P_F\prod_{r=1}^{4}P_{F|F'})=0$).
\end{theorem}

\begin{proof}
We assume a MAP decoder for a fixed sensor network (i.e. fixed and
known $\bFj$'s and $\bxjn$'s);
$g_{\text{MAP}}(\by)=\argmax_jP_{\bFcap|\byn}(\bFj|\by)\propto\argmax_jP_{\byn|\bxn}(\by|\bxj)P_{\bFcap}(\bFj)$.
For this decoder, we consider $P_{e}=\sum_i
P_{e,i}P_{\bFcap}(\bFi)$, where $P_{e,i}$ is averaged over the
random sensor networks. As argued earlier,
$P_{\bFcap}(\bFi)=P_{\bFcap}(\bphi_i)$, and thus we can write
$P_{e}=\sum_{\bphi} P_{e,\bphi}P_{\bFcap}(\bphi)\alpha(\bphi)$
where $\alpha(\bphi)$ corresponds to the number of fields $\bFi$
of field type $\bphi_i$. The quantity
$P_{\bFcap}(\bphi)\alpha(\bphi)$ decays exponentially for
non-typical $\bphi$, and goes to one for the typical $\bphi$, as
$k$ goes to infinity. Thus the average error probability is
dominated by the probability of error for the typical field type
$\bphi^{\ast}$. Note that $P_{\bFcap}(\bphi_i)$ is bounded as
follows,
\begin{equation}
P_{\bFcap}(\bphi_i)\leq
2^{k^2(-D(\bphi_i\|\frac{1}{W}P_F\prod_{r=1}^{4}
P_{F|F'})-H(\bphi_i))} \label{fieldbnd}
\end{equation}
Thus, the typical field type $\bphi^{\ast}$ equals
$\frac{1}{W}P_F\prod_{r=1}^{4} P_{F|F'}$. We bound $P_{e,i}$ for a
field $\bFi$ of typical field type $\bphi^{\ast}$. For large $k$,
this bound will, given the above arguments, bound the average
error probability $P_{e}$.
\begin{equation} \label{Pei}
P_{e,i} =\hspace{-0.8em}
\sum_{\bxi\in\cx^n}\sum_{\by\in\cy^n}P_{\bxn_i}(\bxi)P_{\byn|\bxn}(\by|\bxi)\text{Pr}[\text{error}|i,\bxi,\by]
\end{equation}
To bound \ $\text{Pr}[\text{error}|i,\bxi,\by]$  we define events
$A_{ij}=\{\bxj:P_{\byn|\bxn}(\by|\bxj)P_{\bFcap}(\bFj)\geq
P_{\byn|\bxn}(\by|\bxi)P_{\bFcap}(\bFi) \ | \ i, \bxi,\by \}$.
Since decoding to $j\not\in\cd_i$ results in error,
\begin{align}
\text{Pr}[\text{error}|i,\bxi,\by] & \leq
P(\cup_{j\not\in\cd_i}A_{ij}) \ \ \leq \ \
\sum_{j\not\in\cd_i}P(A_{ij}) \label{errbnd}
\end{align}
We proceed to bound $P(A_{ij})$. For any $s_{ij}\geq0$,
\begin{equation}
P(A_{ij})  =\sum_{\bxj \in A_{ij}}P_{\bxjn|\bxin}(\bxj|\bxi)
\leq\sum_{\bxj\in\cx^n}P_{\bxjn|\bxin}(\bxj|\bxi)\frac{(P_{\byn|\bxn}(\by|\bxj)P_{\bFcap}(\bFj))^{s_{ij}}}{(P_{\byn|\bxn}(\by|\bxi)P_{\bFcap}(\bFi))^{s_{ij}}}
\label{Aijbnd}
\end{equation}
Using (\ref{errbnd}) and (\ref{Aijbnd})  in  (\ref{Pei}),
\begin{equation}  \label{pebnd1}
P_{e,i}\leq \sum_{\bxi\in\cx^n}\sum_{\by\in\cy^n}
P_{\bxn_i}(\bxi)P_{\byn|\bxn}(\by|\bxi)
\sum_{j\not\in\cd_i}\sum_{\bxj\in\cx^n}P_{\bxjn|\bxin}(\bxj|\bxi)\frac{(P_{\byn|\bxn}(\by|\bxj)P_{\bFcap}(\bFj))^{s_{ij}}}{(P_{\byn|\bxn}(\by|\bxi)P_{\bFcap}(\bFi))^{s_{ij}}}
\end{equation}
The bound (\ref{pebnd1}) has an exponential number of terms.
However, it was argued earlier that in our sensor network,
$P_{\bxn_i}(\bxi) = P^{\bgamma_i ,n}(\bx)$ depends only on the
sensor type $\bgamma_i$ of the $i^{th}$ target field, while
$P_{\bxjn|\bxin}(\bxj | \bxi)=P^{\blambda,n}(\bxj | \bxi)$ depends
on the \emph{joint sensor type} $\blambda$ of the $i^{th}$ and
$j^{th}$ target fields. Since we have circular boundary conditions
and $c\geq1$, $\bgamma_i$ and $\bgamma_j$ can be marginalized to
compute $\bphi_i$ and $\bphi_j$ precisely. It was also shown that
$P_{\bFcap}(\bFi)=P_{\bFcap}(\bphi_{i})$. Thus, we can rewrite
(\ref{pebnd1}) by grouping terms according to $\blambda$.
\begin{equation*}
\sum_{j\not\in\cd_i}\sum_{\bxj\in\cx^n}P_{\bxjn|\bxin}(\bxj|\bxi)
\frac{(P_{\byn|\bxn}(\by|\bxj)P_{\bFcap}(\bFj))^{s_{ij}}}{(P_{\byn|\bxn}(\by|\bxi)P_{\bFcap}(\bFi))^{s_{ij}}}=
\end{equation*}
\begin{equation} \label{joint}
\sum_{\blambda \in S_i(D)}\beta(i,\blambda,k)
\sum_{\bxj\in\cx^n}P^{\blambda,n}
(\bxj|\bxi)\frac{(P_{\byn|\bxn}(\by|\bxj)P_{\bFcap}(\bphi_j))^{s_{\blambda}}}{(P_{\byn|\bxn}(\by|\bxi)P_{\bFcap}(\bphi^{\ast}))^{s_{\blambda}}}
\end{equation}
where $S_i(D)$ is the set of joint sensor types that result in an
error. i.e.,
\begin{equation}
S_i(D) = \{\blambda: \blambda \in \Lambda_{k^2}(c),
\lambda_{(0)(1)}+\lambda_{(1)(0)}>D,\; \bgamma_{i,t_1\ldots
t_{|\cs_c|}}= \sum_{\{u_1\ldots
u_{|\cs_c|}\}}\hspace{-1em}\lambda_{(t_1\ldots
t_{|\cs_c|})(u_1\ldots u_{|\cs_c|})}\} \label{sd}
\end{equation}
and where we choose $s_{ij}=s_{\blambda}$ for all $\{i,j\}$ of
joint sensor type ${\blambda}$. Here $\beta(i,\blambda,k)$ is the
number of fields $\bFj$ that have a joint type $\blambda$ with
respect to $\bFi$. $\beta(i,\blambda,k)$ is bounded as,
\begin{eqnarray}
\beta(i,\blambda,k)
\leq \
 2^{k^2(H(\blambda)-H(\bgamma_i ))}
\label{jointcount}
\end{eqnarray}
Combining equations (\ref{pebnd1}),(\ref{joint}),
(\ref{jointcount}), and using the fact that we are bounding a
probability, the following bound holds for
$\rho_{\blambda}\in[0,1]$ and $s_{\blambda}
=\frac{1}{1+\rho_{\blambda}}$.
\begin{multline}
\hspace{-1em}P_{e,i}\leq\hspace{-1em}\sum_{\bxi\in\cx^n}\sum_{\by\in\cy^n}P^{\bgamma_i,n}
(\bxi)P_{\byn|\bxn}(\by|\bxi) \hspace{-1em}\sum_{\blambda \in
S_{\bgamma_i}(D)} \hspace{-1.25em}\big(
2^{k(H(\blambda)-H(\bgamma_i ))}
\sum_{\bxj\in\cx^n}P^{\blambda,n}(\bxj|\bxi)
\frac{(P_{\byn|\bxn}(\by|\bxj)P_{\bFcap}(\bphi_j))^{\frac{1}{1+\rho_{\blambda}}}}{(P_{\byn|\bxn}(\by|\bxi)P_{\bFcap}(\bphi^{\ast}))^{\frac{1}{1+\rho_{\blambda}}}}\big)^{\rho_{\blambda}}
\nonumber
\end{multline}
Using the independence of sensor outputs conditional on the target
vector, the joint p.m.f.s can be simplified as below,
\begin{multline}  \label{pebnd2}
P_{e,i}   \leq  \sum_{\blambda \in S_{\bgamma_i}(D)}
2^{\rho_{\blambda} k^2(H(\blambda)-H(\bgamma_i))}
P_{\bFcap}(\bphi^{\ast})^{\frac{-\rho_{\blambda}}{1+\rho_{\blambda}}}
P_{\bFcap}(\bphi_j)^{\frac{\rho_{\blambda}}{1+\rho_{\blambda}}}
 \Big(\sum_{a_i\in\cx} \sum_{b\in\cy}
P_{Y|X}(b|a_i)^{\frac{1}{1+\rho_{\blambda}}}
\cdot \\
P^{\bgamma_i }(a_i)
(\sum_{a_j\in\cx}P^{\blambda}(a_j|a_i)P_{Y|X}(b|a_j)^{\frac{1}{1+\rho_{\blambda}}})^{\rho_{\blambda}}\Big)^n
\end{multline}
We define the following quantity.
\begin{equation}
E(\rho_{\blambda},\blambda)=
  -\log\Big(\sum_{a_i\in\cx}\sum_{b\in\cy}P^{\bgamma_i }(a_i)
P_{Y|X}(b|a_i)^{\frac{1}{1+\rho_{\blambda}}
}(\sum_{a_j\in\cx}P^{\blambda}(a_j|a_i)P_{Y|X}(b|a_j)^{\frac{1}{1+\rho_{\blambda}}})^{\rho_{\blambda}}\Big)
\end{equation}
Since the number of joint sensor types $\blambda$
 is upper bounded by $(k^2+1)^{|\cs_c|^2}$,
 $k^2=\lceil nR\rceil$, and using (\ref{fieldbnd}), (\ref{pebnd2}) is bounded as,
\begin{multline}
\hspace{-0.5em}P_{e,i}  \leq  2^{-n(-o_1(n)+E_r(R,D))}, E_r(R,D)
\! = \! \min_{\bgamma_i \in T(\bphi^{\ast})} \! \min_{\blambda \in
S_{\bgamma_i}(D)} \max_{0 \leq \rho_{\blambda} \leq 1} \! \!
E(\rho_{\blambda},\blambda)  - \rho_{\blambda}R (H(\blambda)
  -H(\bgamma_i)\\+\frac{1}{1+\rho_{\blambda}}H(\bphi^{\ast})
-\frac{1}{1+\rho_{\blambda}}(D(\bphi_j\|\frac{1}{W}P_F\prod_{r=1}^{4}P_{F|F'})+H(\bphi_j)))
\nonumber
\end{multline}
where $\bgamma_i \in T(\bphi^{\ast})$ consists of the set of
sensor types that marginalize to the typical field type
$\bphi^{\ast}$, and $S_{\bgamma_i}(D)$ is as in (\ref{sd}), with
$\bgamma_i$. Note that $o_1(n) \rightarrow 0$ as $n \rightarrow
\infty$, so we have not included it in the error exponent
$E_r(R,D)$. Observing that $E(0,\blambda)=0 \ \forall \
 \blambda$, we let $\rho_{\blambda}$ go to
zero, rather than optimizing it, thus resulting in a lower bound
on $E_r(R,D)$.
 In the above expression, this implies that in order for $R$
to be achievable
$\frac{E(\rho_{\blambda},\blambda)}{\rho_{\blambda}}-R(H(\blambda)-H(\bgamma_i)+H(\bphi^{\ast})-D(\bphi_j\|\frac{1}{W}P_F\prod_{r=1}^{4}P_{F|F'})-H(\bphi_j))$
must be positive for all $\bgamma,\blambda$, even as
$\rho_{\blambda} \rightarrow 0$.
 But this implies that the derivative of
$E(\rho_{\blambda},\blambda)$ with respect to $\rho_{\blambda}$ at
$\rho_{\blambda}=0$ must be greater than
$R(H(\blambda)-H(\bgamma_i)+H(\bphi^{\ast})-D(\bphi_j\|\frac{1}{W}P_F\prod_{r=1}^{4}P_{F|F'})-H(\bphi_j))$.
It can be easily shown that, $\partial
E(\rho_{\blambda},\blambda)/\partial
\rho_{\blambda}\big|_{\rho_{\blambda}=0}=
D(P_{X_iY}^{\bgamma}\|Q_{X_iY}^{\blambda})$.
Using this derivative in the analysis above, and relaxing
 the conditions $\blambda  \ \in \ {\Lambda}_{k^2}(c)$ by dropping
 the restriction that target fields are restricted to area
 $k^2$
in the definition (\ref{sd}) of $S_{\bgamma_i}(D)$ (thus,
weakening the bound),
 we see that
the sensor network can achieve any rate $R$ bounded as below.
\begin{equation}
R \leq \min_{\bgamma_i \in T(\bphi^{\ast})} \min_{\begin{subarray}{c}\blambda \\
\lambda_{(0)(1)}+\lambda_{(1)(0)}>D
\end{subarray}}\frac{D\left(P_{X_iY}^{\bgamma_i}\|Q_{{X_{i}Y}}^{\blambda}\right)}{DENOM}
\end{equation}
where
$DENOM=H(\blambda)-H(\bgamma_i)+H(\bphi^{\ast})-D(\bphi_j\|\frac{1}{W}P_F\prod_{r=1}^{4}P_{F|F'})-H(\bphi_j)$.
Therefore the Right Hand Side is a lower bound on $C(D)$.
\end{proof}

For the case of $c=0$, the proof has one primary difference. Since
the field type $\bphi$ can be marginalized to compute the sensor
types $\bgamma$, all the target fields are grouped according to
$\bphi$. We let $\bmu$ be the joint field type of target fields
$\bFi$ (with field type $\bphi^{\ast}$) and $\bFj$. Using these
definitions we can write the sensing capacity theorem for the case
of $c=0$ as follows,

\begin{theorem}[Sensing Capacity for pairwise MRF, $c=0$] The
sensing capacity at distortion $D$ for target field distribution
$P_{\bFcap}$ satisfies,
\begin{equation} \label{theorem}
C(D) \geq C_{LB}(D) =
 \min_{\begin{subarray}{c}\bmu \\
\lambda_{(0)(1)}+\lambda_{(1)(0)}>D
\end{subarray}}\frac{D\left(P_{X_iY}^{\bgamma_i}\|Q_{{X_{i}Y}}^{\blambda}\right)}{DENOM}
\end{equation}
where
$DENOM=H(\bmu)-D(\bphi_j\|\frac{1}{W}P_F\prod_{r=1}^{4}P_{F|F'})-H(\bphi_j)$,
where $\bphi^{\ast}$ corresponds to the typical type, and where
$\bphi^{\ast}$, $\bphi_j$,$\bgamma_i$, and $\blambda$ are obtained
by marginalizing the joint field type $\bmu$.
\end{theorem}

\section{Capacity bound examples} \label{sec:sim}

\begin{figure}[t!]
\begin{center}
\includegraphics[height=6cm,keepaspectratio=true]{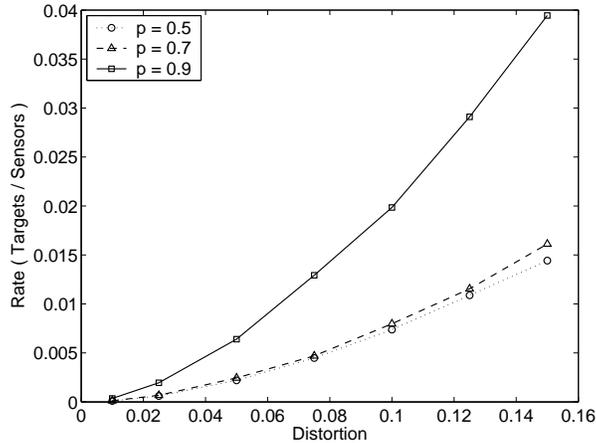}
\vspace{-0.15in} \caption{$C_{LB}(D)$ curves for environments with
different probability distributions (e.g. higher $p$ implies
higher target sparsity).} \label{plot1}
\end{center}
\vspace{-0.27in}
\end{figure}

We compute the capacity bound $C_{LB}(D)$ for environments with
probabilistic models of the form $P_F=[p\;(1-p)]$ and
$P_{F|F'}=[p\; (1-p);\; (1-p)\; p]$ where $p\in[0,1]$. In Figure
\ref{plot1}, we demonstrate the effect of structure in the
environment on $C_{LB}(D)$ by varying $p$. $p=0.5$ corresponds to
an unstructured environment (all $\bF$ equally likely), and
increasing values of $p$ correspond to increasing spatial
structure (e.g. increasing target sparsity). We assume that the
sensors have range $c=0$ (i.e. they sense only one target) and
that the sensing function $\Psi$ is the identity function. The
sensor noise model assumes that the sensor's output is flipped
with probability $0.1$. Figure \ref{plot1} demonstrates that
$C_{LB}(D)$ increases for more structured environments (i.e. fewer
sensors are needed as $p$ increases).

%

\bibliographystyle{IEEEbib}
\bibliography{isit_compact}

\begin{thebibliography}{10}

\bibitem{DuarteMelo03}
E.~J. Duarte-Melo and M.~Liu,
\newblock ``Data-gathering wireless sensor networks: organization and
  capacity,''
\newblock {\em Computer Networks: Special Issue on Wireless Sensor Networks},
  vol. 43, 2003.

\bibitem{Scaglione02}
A.~Scaglione and S.~D. Servetto,
\newblock ``On the interdependence of routing and data compression in multi-hop
  sensor networks,''
\newblock in {\em Proc. 8th ACM Int. Conference on Mobile Computing and
  Networking}, Sept. 2002.

\bibitem{Gatspar04}
M.~Gastpar and M.~Vetterli,
\newblock ``Power-bandwidth-distortion scaling laws for sensor networks,''
\newblock in {\em Third. Int. Symp. Info. Proc. in Sensor Networks}, Apr. 2004.

\bibitem{Li01}
S.Z. Li,
\newblock {\em Markov Random Field Modeling in Image Analysis},
\newblock Springer-Verlag, 2001.

\bibitem{RachlinITW04}
Y.~Rachlin, R.~Negi, and P.~Khosla,
\newblock ``Sensing capacity for target detection,''
\newblock in {\em Proc. IEEE Inform. Theory Wksp.}, Oct. 24-29 2004.

\bibitem{RachlinIPSN05}
Y.~Rachlin, R.~Negi, and P.~Khosla,
\newblock ``Sensing capacity for discrete sensor network applications,''
\newblock in {\em Proc. Fourth Int. Symp. on Inform. Proc. in Sensor Networks},
  April 25-27 2005.

\bibitem{Varshney97}
P.~Varshney,
\newblock {\em Distributed Detection and Data Fusion},
\newblock Springer-Verlag, 1997.

\bibitem{Chakrabarty01}
K.~Chakrabarty, S.~S. Iyengar, H.~Qi, and E.~Cho,
\newblock ``Coding theory framework for target location in distributed sensor
  networks,''
\newblock in {\em Proc. Int. Conf. on Inform. Technology: Coding and
  Computing}, April 2001.

\bibitem{Gallager68}
R.G. Gallager,
\newblock {\em Information Theory and Reliable Communications},
\newblock Wiley, 1968.

\bibitem{Csiszar98}
I.~Csiszar,
\newblock ``The method of types,''
\newblock {\em IEEE Trans. Inform. Theory}, vol. 44, no. 6, 1998.

\end{thebibliography}

\end{document}